\documentclass[letterpaper,9pt,twoside]{article}

\usepackage{mathptmx}
\usepackage[margin=1in]{geometry}
\usepackage{titlesec}
\usepackage{fancyhdr}
\usepackage[hidelinks]{hyperref}
\usepackage[numbers]{natbib}
\usepackage{graphicx}

\fancypagestyle{plain}{
  \fancyhf{}
  
}

\titleformat{\section}{\bfseries\large}{\thesection}{1em}{}

\begin{document}

\thispagestyle{plain}

\begin{flushleft}
{\bfseries\LARGE A Framework for AI-Supported Mediation in Community-based Online Collaboration\par}

\vspace{1em}

{\scshape SOOBIN CHO}, Human Centered Design \& Engineering, University of Washington, USA\\
{\scshape MARK ZACHRY}, Human Centered Design \& Engineering, University of Washington, USA\\
{\scshape DAVID W. MCDONALD}, Human Centered Design \& Engineering, University of Washington, USA
\end{flushleft}

\vspace{1em}

\noindent\textbf{Reference Format:}\\
Soobin Cho, Mark Zachry, and David W. McDonald. 2025. A Framework for AI-Supported Mediation in Community-based Online Collaboration. arXiv:2509.10015 [cs.HC] https://arxiv.org/abs/2509.10015

\vspace{1em}

\section{Introduction: From Moderation to Mediation in Online Collaboration}

Online spaces involve diverse communities engaging in various forms of collaboration, which naturally give rise to discussions, some of which inevitably escalate into conflict or disputes. To address such situations, AI has primarily been used for moderation, such as Reddit’s AutoModerator, which automatically filters content, enforces rules, and detects spam \cite{he2022}. As AI capabilities have advanced, researchers have explored more sophisticated systems to support moderation, such as a visual system that helps Discord moderators monitor multiple channels and detect early signs of toxicity \cite{Choi2023}, and AI models tailored to specific Discord communities that identify user intentions and moderate toxic comments in context \cite{axelsen2023aimoderateonlinecommunities}.

Moderation systems are important because they help maintain order. But common moderation strategies of removing or suppressing content and users rarely address the underlying disagreements or the substantive content of disputes. Moderation systems tend to be reactive and only intervene after visible conflict has already emerged, sometimes after the damage is done. As a result, the underlying conflicts are not guided toward constructive or mutually beneficial outcomes.

Mediation, by contrast, offers a fundamentally different approach. Instead of suppressing disagreement, it fosters understanding, reduces emotional tension, and facilitates consensus through guided negotiation. Mediation not only enhances the quality of collaborative decisions but also strengthens relationships among group members. For this reason, we argue for shifting focus toward AI-supported mediation in community-based online collaboration. We acknowledge that mediation can be time and effort intensive, which motivates the question of whether AI can be leveraged in a mediation system to reduce the overall burden.

Mediating group conflict in online collaboration through AI is a complex problem that is deeply intertwined with group dynamics, processes, and outcomes. In this context, AI no longer acts as a peripheral technical tool, but takes on a more central role within the group. Importantly, such engagement can open up new and more creative forms of collective practice.

\section{Background and Related Work}

\subsection{Definition of Mediation and Mediator Actions}

Mediation, originally rooted in the legal system as a form of alternative dispute resolution, is a facilitated process in which a neutral third party helps disputants voluntarily reach a mutually acceptable agreement without imposing a decision \cite{Lohvinenko2021, Branting2022, Goltsman2009, Jones2000, Janier2015}. Some common styles of mediation include facilitative, evaluative, and transformative mediation, each distinguished by the mediator’s role and approach \cite{Lohvinenko2021, zumeta2000styles}.

While the specifics may vary by context, mediation typically proceeds through six stages: the mediator begins by introducing the process and building trust; the parties then present their perspectives; key issues are identified and structured; possible solutions are generated; these options are evaluated and refined; and finally, the process concludes with agreement or closure \cite{Alfini2013, Moore2014}. Throughout these stages, mediators perform a wide range of actions to guide the dialogue and support resolution. For example, Branting et al. systematically categorized 30 types of mediator speech acts, such as acknowledgement, refocusing, fact-checking, and restating positions \cite{Branting2022}. Similarly, Janier et al. identified six types of mediator tactics, including redirection, clarification of misunderstandings, and addressing negative collateral implications \cite{Janier2015}.

As seen above, existing research on mediation has focused on what mediators do, identifying how the process unfolds and categorizing the concrete actions involved.

\subsection{AI Systems for Online Mediation}

In legal contexts, AI-based mediation systems have primarily appeared in the context of Online Dispute Resolution (ODR). ODR is a form of dispute resolution that leverages the strengths of the Internet, where technology acts as a ``fourth party'' that works alongside the mediator \cite{Rifkin2010}. In ODR, AI performs specific tasks such as analyzing outcomes of similar past cases, drafting legal documents and presentations, generating joint agreements, and assisting in legal research \cite{einy, Abbott}. For example, SAMA in India analyzes case data, predicts outcomes, and suggests possible resolutions \cite{Karthikeyan}; +Acordo, a Brazil-based platform, collects and analyzes claim details using pre-defined parameters to propose settlements \cite{Motta}; and Rechtwijzer in the Netherlands gathers participants’ stance and priorities to identify points of agreement and recommend solutions \cite{Zeleznikow}.

In the HCI domain, researchers have explored how AI systems can support online group discussions as both a mediator and a facilitator. Some systems focus on conversational guidance, as demonstrated by Chen et al.'s study, in which AI steps in during ad hoc team discussions through actions such as asking for clarification, slowing down hasty decisions, restoring speaking turns, and introducing alternative perspectives \cite{Chen}. Others place greater emphasis on participation balancing. Kim et al. and Do et al. developed group chat agents that manage time and actively prompt low‐contributing members to participate \cite{Do1, Do2, Kim1, Kim2}. Finally, AI has also been used to support consensus building. Tessler et al.’s Habermas Machine collected individual opinions from ad hoc groups on political topics and iteratively generated group statements that the group as a whole can agree on \cite{Tessler}.

Across both domains, these systems are primarily designed to assist or automate mediator actions, using participant stance and discussion content as the main, and often only, source of information. Overall, unlike the action side, the informational side of mediation has been largely overlooked.

\section{Suggesting Framework}

The role of the mediator has been described as ``communicating selected information back and forth''  \cite{Singer1990, Goltsman2009}, emphasizing the importance of their access to and understanding of information. However, as discussed earlier, prior work has focused largely on mediator actions, with little attention to the informational side of mediation.

This gap becomes especially problematic in community contexts, where mediation relies on a much more complex and layered understanding of information. In such settings, mediators must go beyond the surface-level content of discussions to grasp deeper contextual knowledge, including group histories and social dynamics. Ultimately, in community-based mediation, it is the depth of what the mediator knows about the group that makes effective mediation possible.

\subsection{Insights from Real-World Cases}

Reflections on real-world cases of group mediation highlight how effective mediation often requires information beyond the discussion content.

For instance, in a PhD milestone exam I attended as a student, one presenter faced a barrage of difficult questions from a faculty member, escalating tension in the room. The professor originally assigned to oversee the session was absent, and a staff member eventually stepped in only to announce that time was up. What was truly needed, however, was a mediator who understood not just the academic content, but the exam's purpose, the student’s level of experience, and the faculty-student dynamic. Without such contextual knowledge, the situation could not be steered constructively, turning what could have been a productive dialogue into a demoralizing experience.

In another case, as a UX researcher I temporarily filled in for a product manager where I led a cross-functional meeting. During the meeting a conflict arose between designers and developers. Although I understood both design and development workflows, I hesitated to intervene because I lacked knowledge of the power dynamics and past practices, such as whose perspectives typically carried more weight or how such disputes were usually resolved. The discussion ended without resolution leaving everyone dissatisfied, and leaving behind unresolved tensions.

These examples illustrate the challenge of effective mediation. Mediation is not simply a process that is executed, it has topical, contextual, and cultural aspects that good mediators either know or can quickly grasp.

\subsection{Information-Focused Framework}

\begin{figure}
    \centering
    \includegraphics[width=0.8\linewidth]{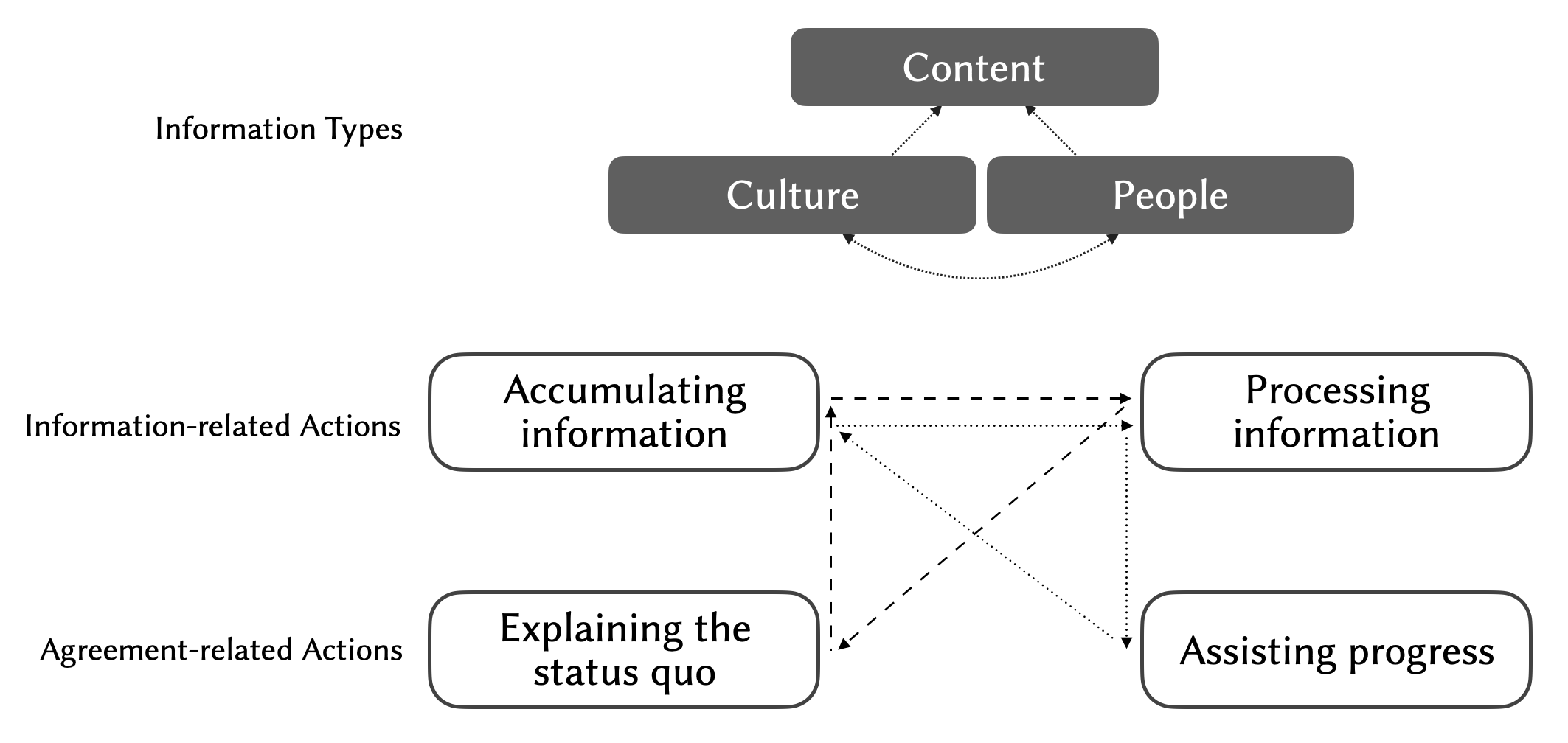}
    \caption{Information-focused framework for AI mediation in community-based collaboration.}
    \label{fig:framework}
\end{figure}

Building on gaps left by prior research, we propose an information-focused framework for AI-supported mediation designed for community-based collaboration (Figure ~\ref{fig:framework}). Within this framework, we hypothesize that AI must acquire and reason over three key types of information: content, culture, and people.

\textit{Content} refers to the topic under discussion and the claims being made. It is the most fundamental element of any discussion. To fully understand the content, however, it is also necessary to be familiar with both the culture and the people involved. \textit{Culture} refers to the shared goals, norms, and expectations that shape how the discussion takes place. \textit{People} refers to the participants’ experiences, roles, and relationships. Importantly, culture and people are mutually dependent: understanding the culture requires knowing who the participants are, and understanding the participants requires awareness of the cultural norms they operate within.

The mediator performs four actions across the three information types, including two information-related actions (accumulating and processing) and two agreement-related actions (explaining the status quo and assisting progress). The mediator operates through two recurring cycles of actions. The first cycle consists of accumulating and processing information to contextually explain the current status, while the second cycle consists of accumulating and processing information to assist progress. These two cycles alternate and together form a larger repetitive mediation process, where the outcome of each cycle becomes the basis for the next one.

\section{Research Design and Trajectory}

Our goal is to assess the feasibility of the proposed framework and demonstrate its potential as a foundation for designing AI mediation systems that support community-based online collaboration. To this end, we explore how the framework’s foundational elements operate in real-world online community settings.

We situate our study in Wikipedia, a well-established online community where collaboration continuously occurs around the creation of encyclopedic content. Wikipedia also has a long-standing and actively updated community culture that is extensively documented. It brings together contributors with diverse experiences and perspectives, whose personal histories and self-introductions are visible through public user pages. At the same time, prior research has noted that the absence of mediation in Wikipedia contributes to two major issues: a decline in the quality of collaborative decisions (i.e., article quality) and the deterioration of relationships among contributors (e.g., members leaving the community) \cite{Jemielniak2014}.

Our research takes a phase-by-phase approach to examine how the proposed mediator actions should operate across the three types of information: content, culture, and people. So far, we have completed a study on how content information should be accumulated and processed, and we are currently investigating how it can be contextually explained.

The first study, accepted to \textit{CSCW 2025}, involved 14 interviews with Wikipedians to understand how they make sense of disputes \cite{Cho}. Each interview consisted of three parts: (1) reading an actual Wikipedia discussion using a think-aloud protocol, (2) writing a summary of the discussion, and (3) reviewing a pre-generated AI summary of the same discussion. The findings revealed 13 key elements that participants considered important for understanding the discussion, such as usernames, evaluation of sources, editor behaviors, and rule violations. Participants noted that the AI summaries often failed to capture community-specific elements among these.

The ongoing study on content explanation involves a web-based prototype that displays a Wikipedia discussion thread with supporting tools. These tools provide various explanations of the discussion content, such as multiple types of summaries and extracted links. Through user interviews, we are examining how Wikipedians use these tools to understand discussions, what they find helpful, and what additional design features they would like to see.

\section{Research Contribution}
Our research clarifies and demonstrates the informational aspects of AI-supported mediation in community settings, which have been largely unaddressed in prior work. By taking into account complex contextual dynamics, our framework brings AI-supported mediation closer to the realities of communities and offers a more holistic perspective. In addition, the framework serves as a foundational guide regardless of the stage or scope of AI, whether it is implemented fully, partially, or in specific components.

\bibliographystyle{ACM-Reference-Format}
\bibliography{references}

\end{document}